\documentclass[letterpaper, 10 pt, conference]{ieeeconf}  
\IEEEoverridecommandlockouts                              
\usepackage{amsmath,amsfonts,amssymb}
\usepackage{tabularx,xtab,array,float}
\usepackage{epsfig,epstopdf,wrapfig,psfrag}
\usepackage{graphicx}
\usepackage{tikz}
\usepackage{balance,physics,stackengine}
\usepackage{cite}
\usepackage[normalem]{ulem}
\usepackage{hyperref}

\makeatletter
\newcommand*{\encircled}[1]{\relax\ifmmode\mathpalette\@encircled@math{#1}\else\@encircled{#1}\fi}
\newcommand*{\@encircled@math}[2]{\@encircled{$\m@th#1#2$}}
\newcommand*{\@encircled}[1]{%
\tikz[baseline,anchor=base]{\node[draw,circle,outer sep=0pt,inner sep=.2ex] {#1};}}
\makeatother

\newtheorem{remark}{Remark}
\newtheorem{theorem}{Theorem}

\newtheorem{corollary}{Corollary}

\newcommand{\qac}{h}
\newcommand{\dqac}{v}
\newcommand{\delths}{{\Delta\theta^*}}
\newcommand{\thetaodd}{\theta^\mathrm{odd}}
\newcommand{\thetaeven}{\theta^\mathrm{even}}
\newcommand{\Hk}{(\eta_{xk} \cot\theta_k + \eta_{yk})}
\newcommand{\drho}{D\rho}

\begin{document}
%
\title{\LARGE \bf Planar Juggling of a Devil-Stick using Discrete VHCs}
%
%
%
\author{Aakash~Khandelwal and~Ranjan~Mukherjee,~\IEEEmembership{Senior Member}
\thanks{This work was supported by the National Science Foundation, Grant CMMI-2043464}%
\thanks{The authors are with the Department of Mechanical Engineering, Michigan State University, East Lansing, MI 48824, USA
{\tt\footnotesize khande10@egr.msu.edu}, {\tt\footnotesize mukherji@egr.msu.edu}}
}
	
\maketitle
\thispagestyle{empty}
\pagestyle{empty}

\begin{abstract}
Planar juggling of a devil-stick using impulsive inputs is addressed using the concept of discrete virtual holonomic constraints (DVHC). The location of the center-of-mass of the devil-stick is specified in terms of its orientation at the discrete instants when impulsive control inputs are applied. The discrete zero dynamics (DZD) resulting from the choice of DVHC provides conditions for stable juggling. A control design that enforces the DVHC and an orbit stabilizing controller are presented. The approach is validated in simulation.
\end{abstract}


\section*{Nomenclature}
\begin{tabularx}{\columnwidth}{lX}
$g$ & acceleration due to gravity, (m/s$^2$)\\
$h_x, h_y$ & Cartesian coordinates of center-of-mass of the devil-stick, (m) \\
$\ell$ & length of the devil-stick, (m)\\
$m$ & mass of the devil-stick, (kg)\\
$r$ & distance of point of application of impulsive force from center-of-mass of the devil-stick, (m) \\
$v_x, v_y$ & Cartesian components of velocity of center-of-mass of the devil-stick, (m/s) \\
$I$ & impulse of impulsive force applied on the devil-stick, (Ns)  \\
$J$ & mass moment of inertia of the devil-stick about its center-of-mass, (kgm$^2$)\\
$\theta$ & orientation of the devil-stick, measured positive counter-clockwise with respect to the horizontal axis, (rad) \\
$\omega$ & angular velocity of the devil-stick, (rad/s) 
\end{tabularx}

\section{Introduction} \label{sec1}

We consider the problem of planar juggling of a devil-stick, which is a nonprehensile manipulation task using impulsive inputs. The dynamics and control of juggling has been considered in \cite{zavala-rio_control_1999, brogliato_control_2000, spong_impact_2001, buehler_planning_1994, sepulchre_stabilization_2003, sanfelice_hybrid_2007}. The combined control problem of the object and robot, with the associated unilateral constraints and impact laws, was considered in \cite{zavala-rio_control_1999} for one-DOF ball juggling, and \cite{brogliato_control_2000} for a class of complementary slackness jugglers. Other strategies for juggling point objects have appeared in \cite{spong_impact_2001, buehler_planning_1994, sepulchre_stabilization_2003, sanfelice_hybrid_2007}.
In contrast to point objects, juggling a devil-stick \cite{schaal_open_1993, kant_non-prehensile_2021, kant_juggling_2022, khandelwal_nonprehensile_2023} requires modeling orientation in addition to position. The dynamics and closed-loop control design for planar \emph{symmetric} juggling appeared in \cite{kant_non-prehensile_2021, kant_juggling_2022}. A coordinate transformation was used to convert the orbital stabilization problem to one of fixed-point stabilization using a Poincar\'e map in a rotating reference frame. Consequently, the dynamic model only permitted orbits with states that are \emph{symmetric} about the vertical axis.\  

We use discrete virtual holonomic constraints (DVHCs), introduced in \cite{khandelwal_discrete_2025}, to define the trajectory of the devil-stick at discrete instants in the inertial frame. Similar to VHCs \cite{maggiore_virtual_2013, mohammadi_dynamic_2018, shiriaev_constructive_2005, grizzle_asymptotically_2001}, DVHCs are geometric constraints on the generalized coordinates that avoid the need for tracking a time-varying reference trajectory, making the approach robust to unmodeled losses. 
However, while VHCs are enforced using continuous feedback such that they are satisfied for all time, DVHCs are enforced using discrete feedback and are satisfied only at the instants when impulsive inputs are applied. The DVHC approach applies to a wide class of underactuated mechanical systems with actuation capabilities restricted to impulsive inputs.
This paper extends the concept of DVHCs from \emph{rotations} to \emph{oscillations} of the passive coordinate for the dynamical system in \cite{khandelwal_discrete_2025}. 
Like the zero dynamics induced by VHCs, the discrete zero dynamics (DZD) induced by DVHCs can be used to infer nonlinear system stability.

The main contributions of this work, in comparison to \cite{kant_non-prehensile_2021}, are highlighted below.
\begin{itemize}
\item The dynamics and trajectory design are performed in the inertial reference frame, eliminating the need for coordinate transformation prior to every control action.
\item The DZD shows that the system permits infinitely many symmetric or asymmetric 2-periodic juggling motions, unlike \cite{kant_non-prehensile_2021, kant_juggling_2022}, which required juggling motions to be symmetric in both position and velocity about the vertical axis.
\end{itemize}
The impulse controlled Poincar\'e map (ICPM) approach \cite{kant_orbital_2020, kant_juggling_2022} is adapted to stabilize any 2-periodic juggling motion. 

This paper is organized as follows. In Section \ref{sec2}, we define the planar devil-stick juggling problem and present the hybrid dynamics of the system. In Section \ref{sec3}, we present the DVHC approach to designing the trajectory of the devil-stick, and a control design that enforces the DVHCs. In Section \ref{sec4}, we derive the DZD, which provides conditions for stable, periodic juggling. In Section \ref{sec5}, we describe a method to stabilize a desired juggling motion. In Section \ref{sec6}, we validate the approach using simulations. The Appendix presents results for the special case when the devil-stick is juggled between orientations that are symmetric about the vertical axis.

\section{System Dynamics} \label{sec2}
\begin{figure}[t!]
    \centering
    \psfrag{A}{$x$}
    \psfrag{B}{$y$}
    \psfrag{C}{$\thetaeven$}
    \psfrag{D}{$\thetaodd$}
    \psfrag{E}{$(h_x, h_y)$}
    \psfrag{F}{$(h_x, h_y)$}
    \psfrag{K}{$g$}
    \psfrag{P}{$I$}
    \psfrag{Q}{$r$}
    \includegraphics[width=0.94\linewidth]{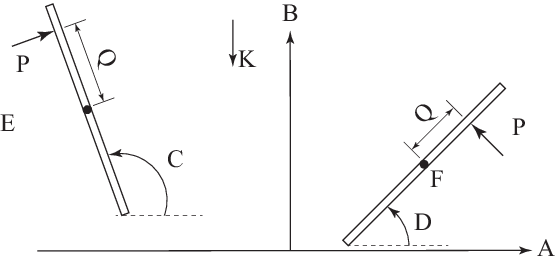}
    \caption{A devil-stick in the vertical plane with configuration variables $(h_x, h_y, \theta)$, and control variables $(I, r)$.}
    \label{fig:description}
\end{figure}

\subsection{System Description} \label{sec21}

Consider the three-DOF devil-stick in the vertical $xy$ plane in Fig. \ref{fig:description}. The devil-stick has length $\ell$, mass $m$, and mass moment of inertia $J$ about its center-of-mass. Its configuration is described by the independent generalized coordinates $(h_x, h_y, \theta)$, where $h_x, h_y$ denote the Cartesian coordinates of its center-of-mass, and $\theta$ denotes its orientation measured positive CCW with respect to the horizontal axis. It is juggled using the control inputs $I$ and $r$, where $I$ is the impulse applied normal to the devil-stick and $r$ is the distance of its point of application from the center-of-mass.
The devil-stick is juggled between the orientations $\thetaodd \in (0, \pi/2)$ and $\thetaeven \in (\pi/2, \pi)$, which are chosen \emph{a priori}. Therefore, $\theta \in [\thetaodd, \thetaeven]$ $\forall t$ during the juggling task.
Impulsive inputs are applied alternately at $\theta = \theta^\mathrm{odd}$ and $\theta = \theta^\mathrm{even}$, and the devil-stick falls freely under gravity between successive applications of impulsive inputs, making the resulting system dynamics hybrid. Since there is one fewer control input than the number of generalized coordinates, the system is underactuated with one passive DOF. The vector of generalized coordinates is
\begin{equation} \label{eq:generalized-coord}
    q \triangleq \begin{bmatrix} h^T & \!\!|\,\,\theta \end{bmatrix}^T, \quad h = \begin{bmatrix} h_x & h_y \end{bmatrix}^T
\end{equation}

\noindent where $h \in \mathbb{R}^2$ are the active coordinates and $\theta \in S^1$ is the passive coordinate. The vector of generalized velocities is
\begin{equation} \label{eq:generalized-vel}
    \dot q \triangleq \begin{bmatrix} v^T & \!\!|\,\,\omega \end{bmatrix}^T, \quad v = \begin{bmatrix} v_x & v_y \end{bmatrix}^T
\end{equation}

\noindent where $v \in \mathbb{R}^2$ and $\omega \in \mathbb{R}$; $v_x, v_y$ denote the Cartesian components of the velocity of the center-of-mass of the devil-stick, and $\omega$ its angular velocity.

\subsection{Hybrid Dynamic Model} 

The hybrid dynamic model captures a single impulsive actuation\footnote{Impulsive inputs cause no change in position coordinates and finite jumps in velocities \cite{kant_orbital_2020}.}, followed by a flight phase under gravity. The dynamic model in \cite{khandelwal_discrete_2025}, which applied to \emph{rotations} of $\theta$, is adapted here for \emph{oscillations} of $\theta$. Using $k$ to represent instants immediately prior to applications of impulsive inputs, the dynamics of the active positions and velocities are \cite{khandelwal_discrete_2025}
\begin{equation} \label{eq:hybrid-pos}
    \qac(k+1) = \qac(k) + \dqac(k) \delta_k 
    + \begin{bmatrix} -\sin\theta_k \\ \cos\theta_k \end{bmatrix} \frac{I_k \delta_k}{m}
    + \begin{bmatrix} 0 \\ - \frac{1}{2}g \delta_k^2 \end{bmatrix}
\end{equation}
\begin{equation} \label{eq:hybrid-vel}
    \dqac(k+1) = \dqac(k)
    + \begin{bmatrix} -\sin\theta_k \\ \cos\theta_k \end{bmatrix} \frac{I_k}{m}
    + \begin{bmatrix} 0 \\ - g \delta_k \end{bmatrix}
\end{equation}

\noindent where $\delta_k$ is the time-of-flight, \emph{i.e.}, the interval between application of the $k$-th and $(k+1)$-th impulsive inputs. Without loss of generality, we assert that the angular configurations $\theta_k$ at which impulsive inputs are applied are such that
\begin{equation} \label{eq:theta-odd-even}
    \theta_k = \begin{cases}
        \thetaodd, \quad\  k\,\ \mathrm{odd} \\
        \thetaeven, \quad\,  k\,\ \mathrm{even}
    \end{cases}
\end{equation}

\noindent The change in orientation between successive applications of the impulsive input is described by
\begin{equation} \label{eq:hybrid-theta-delta}
    \theta_{k+1} - \theta_k = (-1)^{k+1} \delths, \quad \delths \triangleq \thetaeven - \thetaodd 
\end{equation}

\noindent where $\delths > 0$. It immediately follows from \eqref{eq:theta-odd-even} or \eqref{eq:hybrid-theta-delta} that $\theta$ is periodic with period-2, \emph{i.e.},
\begin{equation} \label{eq:theta-periodic}
    \theta_{k+2} = \theta_k
\end{equation}

\noindent The dynamics of the passive states are \cite{khandelwal_discrete_2025}
\begin{align}
    \theta_{k+1} &= \theta_k + \omega_k \delta_k + \frac{I_k r_k}{J} \delta_k \label{eq:hybrid-theta} \\
    \omega_{k+1} &= \omega_k + \frac{I_k r_k}{J} \label{eq:hybrid-omega}
\end{align}

\noindent It follows from \eqref{eq:hybrid-theta-delta}, \eqref{eq:hybrid-theta}, and \eqref{eq:hybrid-omega} that the time-of-flight is
\begin{equation} \label{eq:time-of-flight}
    \delta_k = \frac{(-1)^{k+1} \delths}{\omega_k + \dfrac{I_k r_k}{J}} \quad \Rightarrow \quad \delta_k = \frac{(-1)^{k+1} \delths}{\omega_{k+1}}
\end{equation}

\noindent For feasible planar juggling, $\omega_k < 0$ when $k$ is odd ($\theta_k = \thetaodd$) and $\omega_k > 0$ when $k$ is even ($\theta_k = \thetaeven$). It then follows from the above equation that $\delta_k > 0$ $\forall k$. Further, the control inputs $I_k > 0$, $r_k > 0$ produce a counterclockwise moment when $k$ is odd, and $I_k < 0$, $r_k > 0$ produce a clockwise moment when $k$ is even - see Fig. \ref{fig:description}.

\begin{remark}
    Here, we do not model the robot manipulator; its motion must be controlled to realize the desired control inputs, ensuring that $\begin{bmatrix} \theta - \thetaodd & \thetaeven - \theta \end{bmatrix}^T \geq 0$ $\forall t$ by considering the unilateral constraints between the devil-stick and robot - see the recursive strategy in \cite{zavala-rio_control_1999, brogliato_control_2000}.
\end{remark}

\section{Discrete Virtual Holonomic Constraints} \label{sec3}

\subsection{Constraints on Positions}

We use discrete virtual holonomic constraints (DVHCs) \cite{khandelwal_discrete_2025} to define the trajectory of the center-of-mass of the devil-stick in terms of its orientation at the instants $k$ that impulsive inputs are applied on the devil-stick:
\begin{equation} \label{eq:VHC-rho}
    \rho_k \triangleq \qac(k) - \Phi(\theta_k) = 0, \quad \Phi: S^1 \rightarrow \mathbb{R}^2, \ k = 1,2, \dots 
\end{equation}
\begin{equation} \label{eq:VHC-Phi}
    \Phi(\theta_k) = \begin{bmatrix}
        \Phi_x(\theta_k) &
        \Phi_y(\theta_k)
    \end{bmatrix}^T 
\end{equation}

\noindent where the DVHC components $\Phi_x$, $\Phi_y$ must be well-defined and nonsingular for each $\theta_k \in \{\thetaodd, \thetaeven \}$, and $\Phi_x$ must satisfy $\Phi_x(\thetaodd) > 0$, $\Phi_x(\thetaeven) < 0$.

\subsection{Constraints on Velocities}

We begin by rewriting \eqref{eq:hybrid-pos} as
\begin{equation}
\begin{split}
    \qac(k+1) - \qac(k) = \left\{\dqac(k)
    + \begin{bmatrix} -\sin\theta_k \\ \cos\theta_k \end{bmatrix} \frac{I_k}{m} \right. \\ \left. 
    + \begin{bmatrix} 0 \\ - g \delta_k \end{bmatrix}\right\} \delta_k
    + \begin{bmatrix} 0 \\ \frac{1}{2}g \delta_k^2 \end{bmatrix}
\end{split}
\end{equation}

\noindent Using \eqref{eq:hybrid-vel} in the above equation, we obtain
\begin{equation} \label{eq:hybrid-pos-vel}
    \qac(k+1) - \qac(k) = \dqac(k+1) \delta_k 
    + \begin{bmatrix} 0 \\ \frac{1}{2}g \delta_k^2 \end{bmatrix}
\end{equation}

\noindent Using the DVHC from \eqref{eq:VHC-rho} at $k$ and $k+1$ in the above equation, we obtain
\begin{equation}
    \Phi(\theta_{k+1}) - \Phi(\theta_k) = \dqac(k+1) \delta_k 
    + \begin{bmatrix} 0 \\ \frac{1}{2}g \delta_k^2 \end{bmatrix}
\end{equation}

\noindent The above equation can be rewritten as
\begin{equation}
    \dqac(k+1) = \frac{1}{\delta_k} [\Phi(\theta_{k+1}) - \Phi(\theta_k)] 
    - \begin{bmatrix} 0 \\ \frac{1}{2}g \delta_k \end{bmatrix} 
\end{equation}

\noindent Replacing the instances of $k+1$ by $k$ in the above equation, we obtain
\begin{equation} \label{eq:vhc-vel-time-of-flight}
    \dqac(k) = \frac{1}{\delta_{k-1}} [\Phi(\theta_k) - \Phi(\theta_{k-1})] 
    - \begin{bmatrix} 0 \\ \frac{1}{2}g \delta_{k-1} \end{bmatrix} 
\end{equation}

\noindent From \eqref{eq:theta-periodic} it follows that $\theta_{k-1} = \theta_{k+1}$, and from \eqref{eq:time-of-flight} it follows that $\delta_{k-1} = (-1)^k \delths/\omega_k$. Thus, \eqref{eq:vhc-vel-time-of-flight} can be rewritten as
\begin{equation}
    \dqac(k) = \frac{(-1)^k \omega_k}{\delths} [\Phi(\theta_k) - \Phi(\theta_{k+1})] 
    - \begin{bmatrix} 0 \\ \dfrac{(-1)^k g \delths}{2 \omega_k} \end{bmatrix} 
\end{equation}

The constraint on velocities is therefore given by
\begin{equation} \label{eq:VHC-Drho}
    \drho_k \triangleq \dqac(k) - \Psi(\theta_k, \omega_k) = 0, \quad k = 1,2, \dots 
\end{equation}
\begin{equation} \label{eq:VHC-Psi-Phi}
    \Psi(\theta_k, \omega_k) = 
    \frac{(-1)^k \omega_k}{\delths} [\Phi(\theta_k) - \Phi(\theta_{k+1})] 
    - \begin{bmatrix} 0 \\ \dfrac{(-1)^k g \delths}{2 \omega_k} \end{bmatrix} 
\end{equation}

\begin{remark}
    The value of $\theta_{k+1}$ is known at $k$ from \eqref{eq:hybrid-theta-delta}. In the above equation and subsequent equations, we use $\theta_{k+1}$ for brevity, in place of the full expression $\theta_k + (-1)^{k+1} \delths$ involving $\theta_k$ alone; $\Psi(\theta_k, \omega_k)$ remains a function of only the passive coordinate and passive velocity at $k$.
\end{remark}

\subsection{Control Design Enforcing the DVHC}

We find the control inputs that enforce the DVHC following the approach in \cite{khandelwal_discrete_2025}. Using \eqref{eq:VHC-rho} and \eqref{eq:VHC-Drho} in \eqref{eq:hybrid-pos}, we obtain
\begin{equation} \label{eq:hybrid-pos-rho}
\begin{split}
    \rho_{k+1} + \Phi(\theta_{k+1}) = &[\rho_k + \Phi(\theta_k)] + [\drho_k + \Psi(\theta_k, \omega_k)] \delta_k \\
    &+ \begin{bmatrix} -\sin\theta_k \\ \cos\theta_k \end{bmatrix} \frac{I_k \delta_k}{m}
    + \begin{bmatrix} 0 \\ -\frac{1}{2} g \delta_k^2 \end{bmatrix}
\end{split}
\end{equation}

\noindent Exponential convergence to $\rho_k = 0$ is achieved if 
\begin{equation} \label{eq:rho-lambda}
    \rho_{k+1} = \lambda \rho_k, \quad \lambda \triangleq \mathrm{diag}\begin{bmatrix}
        \lambda_x & \lambda_y
    \end{bmatrix},\, \lambda_x, \lambda_y \in [0,1)
\end{equation}

\noindent We define
\begin{equation} \label{eq:eta-xy}
    \eta_k \triangleq \Phi(\theta_{k+1}) - \Phi(\theta_k), \quad \eta_k = \begin{bmatrix} \eta_{xk} & \eta_{yk} \end{bmatrix}^T
\end{equation}

\noindent and rewrite \eqref{eq:hybrid-pos-rho} as
\begin{equation} \label{eq:control-rho}
\begin{split}
    (\lambda - \mathbb{I}_2) \rho_k + \eta_k = &[\drho_k + \Psi(\theta_k, \omega_k)] \delta_k \\
    &+ \begin{bmatrix} -\sin\theta_k \\ \cos\theta_k \end{bmatrix} \frac{I_k \delta_k}{m}
    + \begin{bmatrix} 0 \\ -\frac{1}{2} g \delta_k^2 \end{bmatrix}
\end{split}
\end{equation}

\noindent The $x$ and $y$ components of \eqref{eq:control-rho} can be expressed separately as
\begin{subequations} \label{eq:control-rho-xy}
\begin{align}
    (\lambda_x-1) \rho_{xk} + \eta_{xk} &- [\drho_{xk} + \Psi_x(\theta_k, \omega_k)] \delta_k \notag \\
    &= - \frac{I_k \delta_k}{m} \sin\theta_k \label{eq:control-rho-x} \\
    (\lambda_y-1) \rho_{yk} + \eta_{yk} &- [\drho_{yk} + \Psi_y(\theta_k, \omega_k)] \delta_k + \frac{1}{2} g \delta_k^2 \notag \\
    &= \frac{I_k \delta_k}{m} \cos\theta_k \label{eq:control-rho-y}
\end{align}
\end{subequations}

\noindent where $\rho_k = \begin{bmatrix} \rho_{xk} & \rho_{yk} \end{bmatrix}^T$, $\drho_k = \begin{bmatrix} \drho_{xk} & \drho_{yk} \end{bmatrix}^T$, and $\Psi = \begin{bmatrix} \Psi_{x} & \Psi_{y} \end{bmatrix}^T$. Since $\sin\theta_k \neq 0$ for any $\theta_k$ in \eqref{eq:theta-odd-even}, we eliminate $I_k$ between \eqref{eq:control-rho-x} and \eqref{eq:control-rho-y} to obtain
\begin{equation} \label{eq:delta-k-quad}
\begin{split}
    \frac{1}{2} g \delta_k^2 &- \left\{[\drho_{xk} + \Psi_x(\theta_k, \omega_k)] \cot\theta_k \right. \\
    &\left. + [\drho_{yk} + \Psi_y(\theta_k, \omega_k)] \right\} \delta_k 
    + \left[\eta_{xk} \cot\theta_k + \eta_{yk} \right. \\
    &\left.+ (\lambda_x-1) \rho_{xk} \cot\theta_k + (\lambda_y-1) \rho_{yk} \right] = 0
\end{split}
\end{equation}

\noindent The time-of-flight $\delta_k$ is obtained by solving the above quadratic equation for the feasible value of $\delta_k$. 
The value of $I_k$ is then obtained from \eqref{eq:control-rho-x} as
\begin{equation} \label{eq:Ik}
    I_k = -\frac{m \left\{ (\lambda_x-1) \rho_{xk} + \eta_{xk} - [\drho_{xk} + \Psi_x(\theta_k, \omega_k)] \delta_k \right\}}{\delta_k \sin\theta_k}
\end{equation}

\noindent Finally, the value of $r_k$ is obtained from \eqref{eq:time-of-flight} as
\begin{equation} \label{eq:rk}
    r_k = \frac{(-1)^{k+1} J \delths}{I_k \delta_k} - \frac{J \omega_k}{I_k}
\end{equation}

\begin{remark}
    The control design must ensure that the time-of-flight $\delta_k > 0$, and the point of application of the impulse lies on the devil-stick, \emph{i.e.}, $r_k \in (-\ell/2, \ell/2)$. 
\end{remark}

\begin{remark}
    It can be shown that \cite{khandelwal_discrete_2025}
    \begin{equation} \label{eq:drho-lambda}
        \drho_{k+1} = \frac{(\lambda - \mathbb{I}_2)\rho_k}{\delta_k}
    \end{equation}
    which ensures that $\drho_k \to 0$ as $\rho_k \to 0$, and the control design presented in this section enforces the DVHC.
\end{remark}
A dead-beat controller arises when $\lambda_x = \lambda_y = 0$; however, this may not be feasible for the robot \cite{brogliato_control_2000}.

If $\rho_k = 0$ and $\drho_k = 0$, explicit solutions of the time-of-flight and control inputs are obtained upon simplification of \eqref{eq:delta-k-quad}, \eqref{eq:Ik} and \eqref{eq:rk} as
\begin{align}
    \delta_k &= \frac{(-1)^{k+1} 2 \omega_k \Hk}{g \delths} \label{eq:delta-k-explicit} \\
    I_k &= \frac{(-1)^{k+1} m \eta_{xk}}{\sin\theta_k} \left[ \frac{\omega_k}{\delths} - \frac{g \delths}{2 \omega_k \Hk} \right] \label{eq:Ik-explicit} \\
    r_k &= - \frac{(-1)^{k+1} J \delths \sin\theta_k}{m \eta_{xk}} \label{eq:rk-explicit}
\end{align}

\section{Discrete Zero Dynamics} \label{sec4}

\subsection{Derivation}

The discrete zero dynamics (DZD) provides the dynamics of passive coordinate $\theta$ and passive velocity $\omega$ when the DVHCs are satisfied, and system trajectories evolve such that $\rho_k \equiv 0 \Rightarrow \qac (k) \equiv \Phi(\theta_k)$ and $\drho_k \equiv 0 \Rightarrow \dqac (k) \equiv \Psi(\theta_k, \omega_k)$.
\noindent It follows from \eqref{eq:VHC-Drho}, \eqref{eq:VHC-Psi-Phi}, and \eqref{eq:eta-xy} that
\begin{subequations} \label{eq:VHC-Psi-k}
\begin{align} 
    v_x (k) &= - \frac{(-1)^k \omega_k \eta_{xk}}{\delths} \\
    v_y (k) &= - \frac{(-1)^k \omega_k \eta_{yk}}{\delths} - \dfrac{(-1)^k g \delths}{2 \omega_k}
\end{align}
\end{subequations}

\noindent Similarly, it follows from \eqref{eq:theta-periodic}, \eqref{eq:VHC-Drho}, \eqref{eq:VHC-Psi-Phi}, and \eqref{eq:eta-xy} that
\begin{subequations} \label{eq:VHC-Psi-kp1}
\begin{align}
    v_x (k+1) &= - \frac{(-1)^k \omega_{k+1} \eta_{xk}}{\delths}  \\
    v_y (k+1) &= - \frac{(-1)^k \omega_{k+1} \eta_{yk}}{\delths} + \dfrac{(-1)^k g \delths}{2 \omega_{k+1}}
\end{align}
\end{subequations}

\noindent since $\eta_{k+1} = \Phi(\theta_{k+2}) - \Phi(\theta_{k+1}) = -[\Phi(\theta_{k+1}) - \Phi(\theta_k)] = -\eta_k$.
The input $I_k$ can be eliminated between the two equations in \eqref{eq:hybrid-vel} to obtain
\begin{equation} 
\begin{split}
    &v_x(k+1) \cos\theta_k + v_y(k+1) \sin\theta_k \\= &v_x(k) \cos\theta_k + v_y(k) \sin\theta_k - g \delta_k \sin\theta_k
\end{split}
\end{equation}

\noindent Using \eqref{eq:time-of-flight}, the above equation may be rewritten as
\begin{equation} \label{eq:eliminate-Ik}
\begin{split}
    &v_x(k+1) \cos\theta_k + v_y(k+1) \sin\theta_k \\= &v_x(k) \cos\theta_k + v_y(k) \sin\theta_k + \frac{(-1)^{k} g \delths}{\omega_{k+1}} \sin\theta_k
\end{split}
\end{equation}

\noindent Using \eqref{eq:VHC-Psi-k} and \eqref{eq:VHC-Psi-kp1} in \eqref{eq:eliminate-Ik} and simplifying, we obtain
\begin{equation} \label{eq:omkp1-zero-dyn}
    \eta_{xk} \cot\theta_k + \eta_{yk} - \frac{g \delths^2}{2 \omega_k \omega_{k+1}} = 0
\end{equation}

\noindent Equations \eqref{eq:hybrid-theta-delta} and \eqref{eq:omkp1-zero-dyn} together define the DZD:
\begin{align} 
&\mathcal{Z}(\theta_k, \omega_k, \theta_{k+1}, \omega_{k+1}) = 0 \notag \\
&\mathcal{Z} \triangleq \begin{bmatrix} \theta_{k+1} - \theta_k - (-1)^{k+1} \delths \\
\eta_{xk} \cot\theta_k + \eta_{yk} - \dfrac{g \delths^2}{2 \omega_k \omega_{k+1}} \end{bmatrix} \label{eq:zerodyn}
\end{align}

\noindent and describe the system behavior when the DVHC in \eqref{eq:VHC-rho} is identically satisfied. The nature of solutions to \eqref{eq:zerodyn} governs the stability and periodicity of juggling. Conditions for stable, periodic juggling are presented next.

\subsection{Stable Juggling}

\begin{theorem} \label{theorem:periodic}
    The dynamical system in \eqref{eq:zerodyn} is periodic with period-2, \emph{i.e.}, $\theta_{k+2} = \theta_k$ and $\omega_{k+2} = \omega_k$, if and only if
    \begin{equation} \label{eq:theta-odd-even-symmetric}
        \Phi_y(\theta_k) = \beta - \sigma \Phi_x(\theta_k), \quad \sigma \triangleq \frac{1}{2}(\cot\thetaodd + \cot\thetaeven)
    \end{equation}
where $\sigma$ is a constant, and $\beta \in \mathbb{R}$ is an arbitrary constant.
\end{theorem}

\begin{proof}
The 2-periodicity of $\theta$ follows directly from \eqref{eq:theta-periodic}. To establish 2-periodicity of $\omega$, observe that \eqref{eq:omkp1-zero-dyn} can be rewritten 
\begin{equation} \label{eq:omk-omkp1}
    \frac{g \delths^2}{2 \omega_k \omega_{k+1}} = \eta_{xk} \cot\theta_k + \eta_{yk}
\end{equation}

\noindent It follows that
\begin{equation} \label{eq:omkp1-omkp2}
    \frac{g \delths^2}{2 \omega_{k+1} \omega_{k+2}} = -\eta_{xk} \cot\theta_{k+1} - \eta_{yk}
\end{equation}

\noindent since $\eta_{k+1} = -\eta_k$ - see the text after \eqref{eq:VHC-Psi-kp1}. Dividing \eqref{eq:omk-omkp1} by \eqref{eq:omkp1-omkp2} and simplifying, we obtain
\begin{equation} \label{eq:omk-omkp2}
    \omega_{k+2} = {- \left( \frac{\eta_{xk} \cot\theta_k + \eta_{yk}}{\eta_{xk} \cot\theta_{k+1} + \eta_{yk}} \right) } \omega_k
\end{equation}

\noindent It follows that
\begin{align*}
    &\omega_{k+2} = \omega_k  \\
    \iff& \eta_{xk} \cot\theta_k + \eta_{yk} = - \eta_{xk} \cot\theta_{k+1} - \eta_{yk} \\
    \iff& \frac{1}{2}(\cot\theta_k + \cot\theta_{k+1}) \eta_{xk} + \eta_{yk} = 0
\end{align*}
Since $\theta_k = \thetaodd \Rightarrow \theta_{k+1} = \thetaeven$, and $\theta_k = \thetaeven \Rightarrow \theta_{k+1} = \thetaodd$, $\frac{1}{2}(\cot\theta_k + \cot\theta_{k+1}) = \sigma\, \forall k$, with $\sigma$ defined in \eqref{eq:theta-odd-even-symmetric}. Thus,
\begin{align*}
    &\omega_{k+2} = \omega_k \\
    \iff& \sigma \eta_{xk} + \eta_{yk} = 0 \\
    \iff& \sigma \Phi_x(\theta_{k+1}) + \Phi_y(\theta_{k+1}) = \sigma \Phi_x(\theta_k) + \Phi_y(\theta_k) \\
    \iff& \sigma \Phi_x(\theta_k) + \Phi_y(\theta_k) = \beta \quad \forall k
\end{align*}
where $\beta$ is a constant. This completes the proof.
\end{proof}

\begin{corollary} \label{cor1}
If $\thetaeven = \pi - \thetaodd$, \emph{i.e.}, the devil-stick is juggled between orientations symmetric about the vertical axis, the DVHC \eqref{eq:VHC-Phi} satisfies $\Phi_y(\theta_k) = \beta$.
\end{corollary}

\begin{proof}
The proof follows directly from \eqref{eq:theta-odd-even-symmetric}, since $\thetaeven = \pi - \thetaodd \Rightarrow \sigma = 0$.
\end{proof}

Theorem \ref{theorem:periodic} implies that a unique stable, 2-periodic juggling motion corresponds to two unique $\omega_k$ values for each $\theta_k \in \{\thetaodd, \thetaeven \}$, when the DVHC satisfies \eqref{eq:theta-odd-even-symmetric}. These values of $\omega_k$ are:
\begin{align} 
    \omega_k &= \begin{cases}
        \omega^*, \quad\quad\ \ k\,\ \mathrm{odd} \\
        \dfrac{g \delths^2}{2 \omega^* \xi}, \quad  k\,\ \mathrm{even}
    \end{cases} \label{eq:omega-odd-even} \\
    \xi &\triangleq -\frac{1}{2} (\cot\thetaodd - \cot\thetaeven) \eta_x^* \\
    \eta_x^* &\triangleq \Phi_x(\thetaodd) - \Phi_x(\thetaeven)
\end{align}

\noindent where $\xi < 0$, $\eta_x^* > 0$ are constants, and $\omega^* < 0$ may be chosen arbitrarily; the value of $\omega_k$ corresponding to even $k$ is found from \eqref{eq:omkp1-zero-dyn} and \eqref{eq:theta-odd-even-symmetric}. Further, it follows from \eqref{eq:delta-k-explicit} and \eqref{eq:omega-odd-even} that there exist the two unique times-of-flight
\begin{equation} \label{eq:delta-k-odd-even}
    \delta_k = \begin{cases}
        {2 \omega^* \xi} / {(g \delths)}, \quad k\,\ \mathrm{odd} \\
        - {\delths} / {\omega^*}, \qquad\ \, k\,\ \mathrm{even}
    \end{cases}
\end{equation}

\noindent From \eqref{eq:Ik-explicit}, \eqref{eq:theta-odd-even-symmetric}, and \eqref{eq:omega-odd-even}, we obtain
\begin{equation} \label{eq:Ik-stable}
    I_k = \frac{(-1)^k m \eta_x^*}{\sin\theta_k} \left[ \frac{\omega^*}{\delths} - \frac{g \delths}{2 \omega^* \xi} \right]
\end{equation}

\noindent From \eqref{eq:rk-explicit} and \eqref{eq:theta-odd-even-symmetric}, we obtain
\begin{equation} \label{eq:rk-stable}
    r_k = \frac{J \delths \sin\theta_k}{m \eta_x^*}
\end{equation}

\noindent These results demonstrate that the DZD \eqref{eq:zerodyn} permits infinitely many stable, 2-periodic orbits between arbitrary $\thetaodd \in (0, \pi/2)$, $\thetaeven \in (\pi/2,\pi)$ when the DVHC \eqref{eq:VHC-Phi} is chosen such that \eqref{eq:theta-odd-even-symmetric} holds, with each orbit corresponding to a unique choice of $\omega^* < 0$.

\begin{remark} \label{rem:identical}
    Identical trajectories of the center-of-mass between successive impacts are obtained when $\delta_k$ is constant $\forall k$ in \eqref{eq:delta-k-odd-even}, \emph{i.e.}, ${2 \omega^* \xi} / {(g \delths)} = - {\delths} / {\omega^*}$, which implies 
    \begin{equation} \label{eq:omega-star-identical}
         \omega^* = -\delths \sqrt{-\frac{g}{2 \xi}} \ \Rightarrow \  
        \omega_k = \begin{cases}
        \ \ \omega^*, \quad k\,\ \mathrm{odd} \\
        -\omega^*, \quad  k\,\ \mathrm{even}
    \end{cases}
    \end{equation}
\end{remark}

\section{Stabilization of Juggling Motion} \label{sec5}

\subsection{Orbit Selection}

A distinct stable juggling motion described by DVHCs satisfying \eqref{eq:theta-odd-even-symmetric} is the orbit\footnote{The orbit can equivalently be specified based on $\omega_k$ $\forall\, \theta_k = \thetaeven$.}
\begin{equation} \label{eq:orbit}
\begin{split}
    \mathcal{O}^* = \{(q, \dot q) : \qac(k) &= \Phi(\theta_k), \dqac(k) = \Psi(\theta_k, \omega_k), \\
    \omega_k &= \omega^* \,\,\forall\, \theta_k  = \thetaodd \}
\end{split}
\end{equation}

\noindent To stabilize $\mathcal{O}^*$, we define the section\footnote{The section can alternatively be chosen with $\theta = \thetaeven,\ \omega > 0$.}
\begin{equation} \label{eq:poincare-section}
    \Sigma = \{(q, \dot q) \in \mathbb{R}^2 \times S^1 \times \mathbb{R}^3 : \theta = \thetaodd,\ \omega < 0\}
\end{equation}

\noindent on which the states are
\begin{equation}
    z = \begin{bmatrix} \qac^T & \dqac^T &\omega \end{bmatrix}^T, \quad z \in \mathbb{R}^5
\end{equation}

\noindent For a system trajectory not on $\mathcal{O}^*$, we modify the inputs $I$ and $r$ only at $\theta_k = \thetaodd$ from the values in \eqref{eq:Ik} and \eqref{eq:rk} such that the system trajectory is driven to $\mathcal{O}^*$. Since $\theta_{k+2} = \theta_k$, the hybrid dynamics of the system between successive intersections with $\Sigma$ (involving two applications of the impulsive inputs) is given by
\begin{equation} \label{eq:hybrid-map}
    z(j+1) = \mathbb{P} [z(j), I(j), r(j)]
\end{equation}

\noindent where $z(j)$ denotes the states on $\Sigma$ immediately prior to application of inputs $I(j)$ and $r(j)$ on $\Sigma$. Note that for every intersection of the system trajectory with $\Sigma$, the increment in $k$ is $2$ while the increment in $j$ is $1$. 

\subsection{Orbital Stabilization}

We recall the approach to orbit stabilization in \cite{khandelwal_discrete_2025}. If $(q, \dot q) \in \mathcal{O}^*$, the system trajectory lies on $\mathcal{O}^*$ under the inputs from \eqref{eq:Ik} and \eqref{eq:rk} alone. Therefore, the intersection of $\mathcal{O}^*$ with $\Sigma$ is a fixed point $z(j) = z^*$, $I(j) = I^*$, $r(j) = r^*$ of the map $\mathbb{P}$; 
$I^*$ and $r^*$ are obtained from \eqref{eq:Ik-stable} and \eqref{eq:rk-stable}:
\begin{equation} \label{eq:fixed-point}
    z^* = \mathbb{P} (z^*, I^*, r^*)
\end{equation}

\noindent If $(q, \dot q) \notin \mathcal{O}^*$, the inputs \eqref{eq:Ik} and \eqref{eq:rk} drive system trajectories to an orbit such that $\rho_k = 0$, but not necessarily the desired orbit $\mathcal{O}^*$. We stabilize $\mathcal{O}^*$ by using the ICPM approach \cite{kant_orbital_2020} to stabilize the fixed point $z^*$ on $\Sigma$. First, linearize $\mathbb{P}$ about $z(j) = z^*$ and $I(j) = I^*,\, r(j) = r^*$ as
\begin{align} 
    e(j+1) &= \mathcal{A} e(j) + \mathcal{B} u(j) \label{eq:linearized-map} \\ 
    e(j) &\triangleq z(j) - z^*, \quad 
    u(j) \triangleq \begin{bmatrix} I(j) \\ r(j) \end{bmatrix} - \begin{bmatrix} I^* \\ r^* \end{bmatrix}
\end{align} 

\noindent where $\mathcal{A} \in \mathbb{R}^{5 \times 5}$, $\mathcal{B} \in \mathbb{R}^{5 \times 2}$ may be computed numerically \cite{khandelwal_discrete_2025}. If the pair $(\mathcal{A}, \mathcal{B})$ is controllable, $\mathcal{O}^*$ is asymptotically stabilized by the discrete feedback
\begin{equation} \label{eq:discrete-feedback}
    u(j) = \mathcal{K} e(j)
\end{equation}

\noindent where $\mathcal{K}$ places the eigenvalues of $(\mathcal{A} + \mathcal{B}\mathcal{K})$ within the unit circle. Thus, for every intersection of the system trajectory with $\Sigma$, the applied inputs are
\begin{equation} \label{eq:orbit-I-r}
    \begin{bmatrix} I(j) \\ r(j) \end{bmatrix} = \begin{bmatrix} I_k \\ r_k \end{bmatrix} + u(j)
\end{equation}

\noindent where $I_k$ and $r_k$ are given by \eqref{eq:Ik} and \eqref{eq:rk}. The additional input $u(j)$ is not applied if the trajectory is sufficiently close to $\mathcal{O}^*$, \emph{i.e.}, $\norm{e(j)}$ is sufficiently small.

\begin{remark}
    Stabilization of a periodic orbit for initial conditions outside the domain of linearization of the desired orbit $\mathcal{O}^*$ may be realized by choosing a sequence of fixed points $z^* \equiv z^*(j)$ on $\Sigma$, and recomputing $\mathcal{A}$, $\mathcal{B}$, and $\mathcal{K}$ to gradually guide the system trajectory to $\mathcal{O}^*$ - see \cite{khandelwal_maneuvering_2024}.
\end{remark}

\section{Simulation} \label{sec6}

The physical parameters of the devil-stick in SI units are:
\begin{equation*}
    m = 0.1, \quad \ell = 0.5, \quad J = \frac{1}{12} m \ell^2 = 0.0021
\end{equation*}

\subsection{Enforcing the DVHC}

\begin{figure}[t]
    \centering
    \psfrag{A}{\hspace{-10pt} \footnotesize{$\theta_k$ (rad)}}
    \psfrag{B}{\hspace{-18pt} \footnotesize{$\omega_k$ (rad/s)}}
    \psfrag{T}{\footnotesize{$k$}}
    \psfrag{Q}{\hspace{-10pt} \footnotesize{$\delta_k$ (s)}}
    \psfrag{F}{\hspace{-10pt} \footnotesize{$I_k$ (Ns)}}
    \psfrag{R}{\hspace{-10pt} \footnotesize{$r_k$ (m)}}
    \psfrag{C}{$\rho_x$}
    \psfrag{D}{$\rho_y$}
    \psfrag{G}{$\drho_x$}
    \psfrag{L}{$\drho_y$}
    \includegraphics[width=\linewidth]{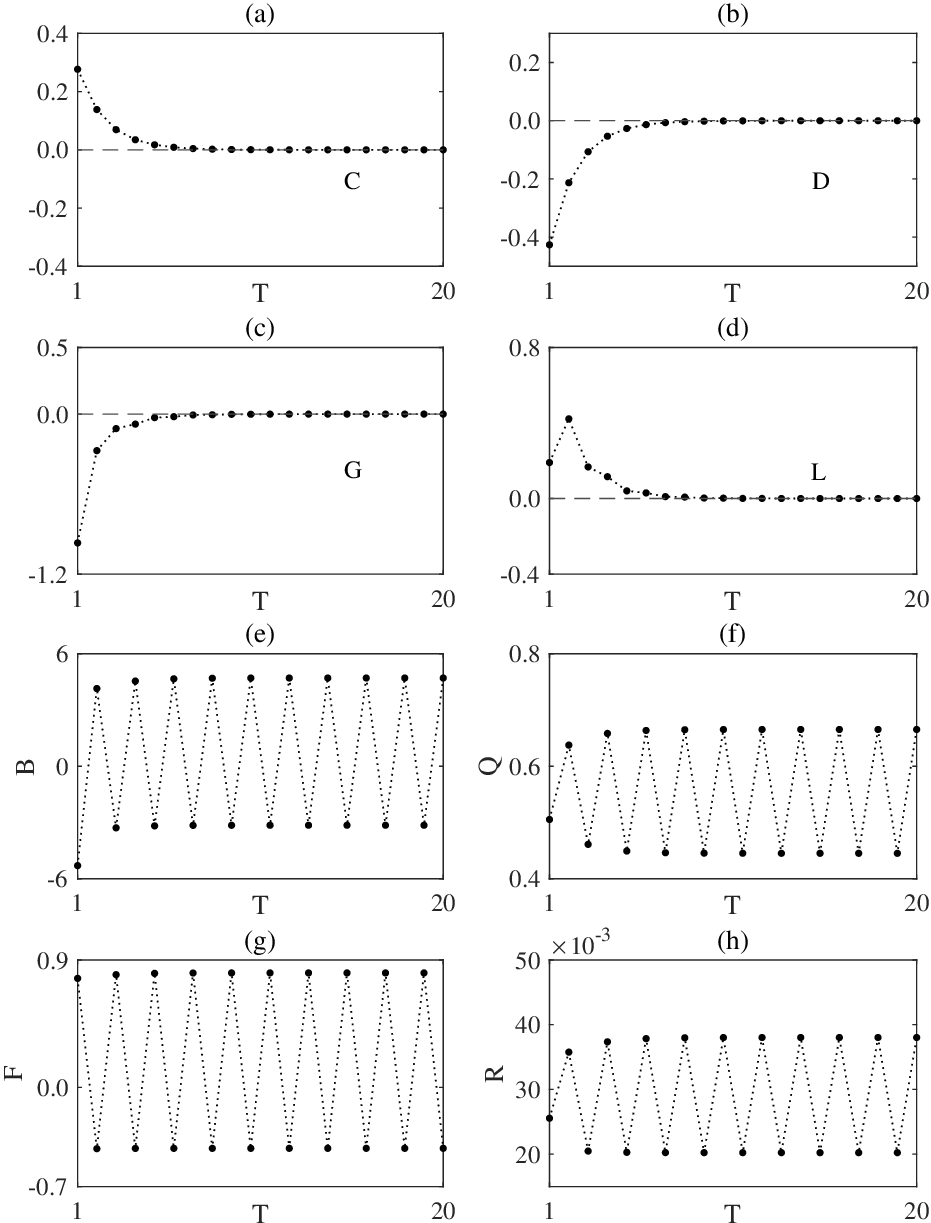}
    \caption{Stabilization of 2-periodic juggling of a devil-stick from arbitrary initial conditions:
    (a)-(b) show the components of $\rho_k$, (c)-(d) show the components of $\drho_k$,
    (e) shows the pre-impact angular velocity $\omega_k$, (f) shows the time-of-flight $\delta_k$
    (g) shows the applied impulse $I_k$, and (h) shows the point of application $r_k$ of the impulsive force.
    }
    \label{fig:sim-vhc}
\end{figure}

We choose a DVHC \eqref{eq:VHC-Phi} of the form
\begin{equation} \label{eq:vhc-sim}
    \Phi(\theta_k) = \begin{bmatrix}
        \alpha\tan\theta_k \\
        \beta - \sigma\alpha\tan\theta_k
    \end{bmatrix}, \quad\alpha = 0.6131,\,\, \beta = 3
\end{equation}

\noindent The values of $\thetaodd$ and $\thetaeven$ are chosen to be
\begin{equation} \label{eq:theta-sim}
    \thetaodd = \pi/9,\,\, \thetaeven = 7\pi/9 \ \Rightarrow\ \delths = 2\pi/3,\,\, \sigma = 0.7779
\end{equation}

\noindent The choices in \eqref{eq:vhc-sim} and \eqref{eq:theta-sim} ensure that the DZD in \eqref{eq:zerodyn} permits stable, 2-periodic orbits - see Theorem \ref{theorem:periodic}. The control inputs are obtained using \eqref{eq:Ik} and \eqref{eq:rk} with
\begin{equation} \label{eq:lambda-sim}
    \lambda_x = \lambda_y = 0.5
\end{equation}

\begin{figure}[b]
    \centering
    \psfrag{X}{\hspace{-10pt} \footnotesize{$h_x$ (m)}}
    \psfrag{Y}{\hspace{-10pt} \footnotesize{$h_y$ (m)}}
    \includegraphics[width=\linewidth]{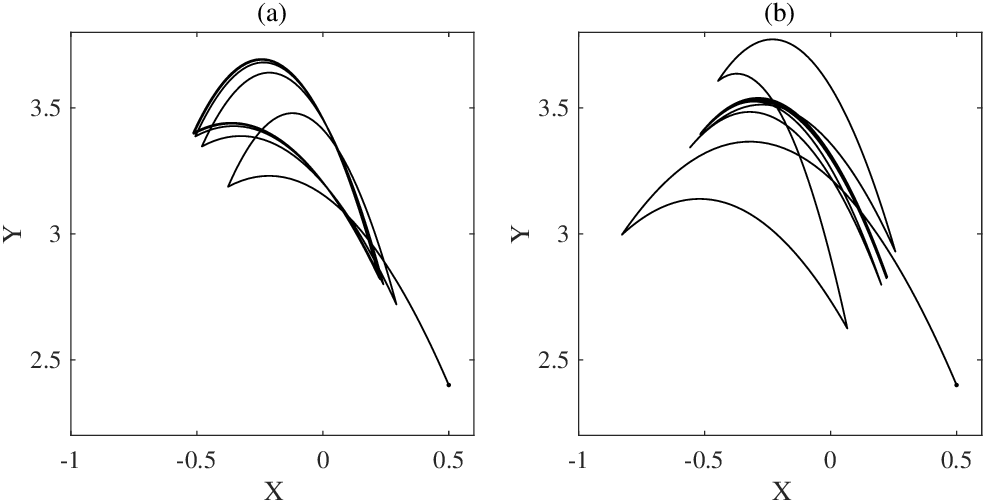}
    \caption{Trajectory of the center-of-mass of the devil-stick corresponding to the results in (a) Fig. \ref{fig:sim-vhc} and (b) Fig. \ref{fig:sim-orbit}.}
    \label{fig:trajectory}
\end{figure}

We consider the initial conditions
\begin{equation} \label{eq:initial-conditions}
    \begin{bmatrix} q^T & \dot q^T \end{bmatrix}^T =
    \begin{bmatrix} 0.5 & 2.4 & 0.3491 & 0.9 & -3.2 & -5.3 \end{bmatrix}^T
\end{equation}

\noindent for which $\theta = \theta_1 = \thetaodd$ and $\rho_1 \neq 0$. The simulation results are shown in Fig. \ref{fig:sim-vhc} for $k = 1$ through $k = 20$, corresponding to a duration of approximately $10.48$ s. The components of $\rho_k$ are shown in Fig. \ref{fig:sim-vhc}(a)-(b), and the components of $\drho_k$ are shown in Fig. \ref{fig:sim-vhc}(c)-(d). The plots demonstrate that the system trajectory converges to $\rho_k = 0$ exponentially. As the DVHC is enforced, the system settles to a stable, 2-periodic juggling motion with 
\begin{equation} \label{eq:periodic-juggling-sim}
    \omega_k = \begin{cases}
        -3.1479,\  k\,\,\mathrm{odd} \\
        \ \ 4.7045,\ k\,\,\mathrm{even}
    \end{cases}
    \!\!\!\!\!\!,\,
    \delta_k = \begin{cases}
        0.4452,\  k\,\,\mathrm{odd} \\
        0.6653,\ k\,\,\mathrm{even}
    \end{cases}
\end{equation}

\noindent which agree with \eqref{eq:omega-odd-even} and \eqref{eq:delta-k-odd-even}. The values of $\omega_k$ and $\delta_k$ are plotted in Fig. \ref{fig:sim-vhc}(e) and Fig. \ref{fig:sim-vhc}(f).
The control inputs $I_k$ and $r_k$ are shown in Fig. \ref{fig:sim-vhc}(g) and Fig. \ref{fig:sim-vhc}(h) respectively. For the 2-periodic juggling motion in \eqref{eq:periodic-juggling-sim}, $I_k = 0.8086$ Ns, $r_k = 0.0202$ m for $k$ odd, and $I_k = -0.4302$ Ns, $r_k = 0.0380$ m for $k$ even, in accordance with \eqref{eq:Ik-stable} and \eqref{eq:rk-stable}. The trajectory of the center-of-mass of the devil-stick corresponding to these results is shown in Fig. \ref{fig:trajectory}(a).

\subsection{Stabilization of a Periodic Orbit}

\begin{figure}[t]
    \centering
    \psfrag{A}{\hspace{-10pt} \footnotesize{$\theta_k$ (rad)}}
    \psfrag{B}{\hspace{-18pt} \footnotesize{$\omega_k$ (rad/s)}}
    \psfrag{T}{\footnotesize{$k$}}
    \psfrag{Q}{\hspace{-10pt} \footnotesize{$\delta_k$ (s)}}
    \psfrag{F}{\hspace{-10pt} \footnotesize{$I_k$ (Ns)}}
    \psfrag{R}{\hspace{-10pt} \footnotesize{$r_k$ (m)}}
    \psfrag{C}{$\rho_x$}
    \psfrag{D}{$\rho_y$}
    \psfrag{G}{$\drho_x$}
    \psfrag{L}{$\drho_y$}
    \includegraphics[width=\linewidth]{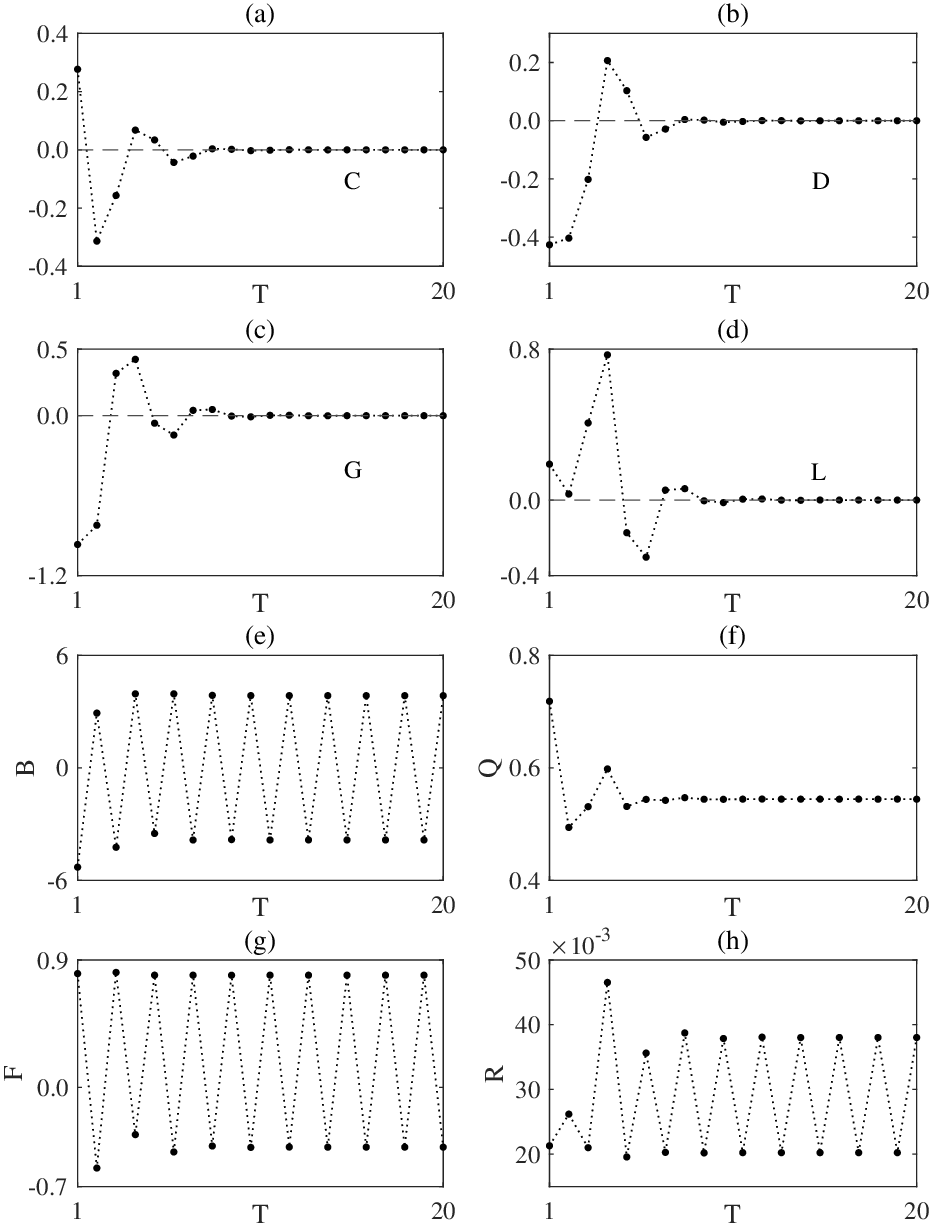}
    \caption{Orbital stabilization of 2-periodic juggling of a devil-stick from arbitrary initial conditions; 
    $\delta_k = 0.5442$ s $\forall k$ corresponding to stable juggling:
    (a)-(b) show the components of $\rho_k$, (c)-(d) show the components of $\drho_k$,
    (e) shows the pre-impact angular velocity $\omega_k$, (f) shows the time-of-flight $\delta_k$
    (g) shows the applied impulse $I_k$, and (h) shows the point of application $r_k$ of the impulsive force.
    }
    \label{fig:sim-orbit}
\end{figure}

We now demonstrate the effectiveness of the controller described in Section \ref{sec5} in stabilizing a desired orbit.
We consider the DVHC defined by \eqref{eq:vhc-sim} and \eqref{eq:theta-sim}. Of the infinitely many stable, 2-periodic orbits, we stabilize the orbit for which $\omega^* = -3.8483$ rad/s, obtained from \eqref{eq:omega-star-identical}. This results in $\delta_k = 0.5442$ s $\forall k$ corresponding to stable juggling. This orbit can be described using \eqref{eq:orbit} as
\begin{equation} \label{eq:orbit-sim}
\begin{split}
    \mathcal{O}^* = \{(q, \dot q) : \qac(k) &= \Phi(\theta_k), \dqac(k) = \Psi(\theta_k, \omega_k), \\
    \omega_k &= -3.8483 \,\,\forall\, \theta_k = \pi/9 \}
\end{split}
\end{equation}

\noindent We obtain the control inputs \eqref{eq:Ik} and \eqref{eq:rk} with the same choice of $\lambda$ as in \eqref{eq:lambda-sim}. The section \eqref{eq:poincare-section} on which the orbit stabilizing controller \eqref{eq:orbit-I-r} acts is
\begin{equation} \label{eq:poincare-section-sim}
    \Sigma = \{(q, \dot q) \in \mathbb{R}^2 \times S^1 \times \mathbb{R}^3 : \theta = \pi/9, \omega < 0\}
\end{equation}

\noindent The fixed point of the map $\mathbb{P}$ in \eqref{eq:fixed-point} corresponding to the orbit $\mathcal{O}^*$ in \eqref{eq:orbit-sim} and the section $\Sigma$ in \eqref{eq:poincare-section-sim} is
\begin{align}
    z^* &= \begin{bmatrix}
        0.2232 & 2.8264 & 1.3553 & -3.7238 & -3.8483
    \end{bmatrix}^T \notag \\
    I^* &= 0.7926, \quad r^* = 0.0202
\end{align}

\noindent The matrices $\mathcal{A}$ and $\mathcal{B}$ in \eqref{eq:linearized-map} are obtained numerically as
\begin{equation*}
\begin{split}
    \mathcal{A} &= \begin{bmatrix}
    0.2500  & -0.0000  & -0.0000  & -0.0000  & -0.0000 \\
    0.0000  &  0.2500  &  0.0000  & -0.0000  &  0.0000 \\
    0.0422  &  0.3498  &  0.6976  &  0.2539  & -0.0000 \\
    0.5976  & -0.0428  &  0.8307  &  0.3025  & -0.0000 \\
   -1.4247  & -0.9931  & -1.9809  & -0.7209  &  0.0000
    \end{bmatrix}
    \\ 
    \mathcal{B} &= \begin{bmatrix}
    0.0000  &  3.6387 &  -8.4460 & -16.8780 &  23.9812 \\
   30.4410  & 26.3808 & -147.1159 & -193.4499 & 282.1914
    \end{bmatrix}^T
\end{split}
\end{equation*}

\noindent While not all eigenvalues of $\mathcal{A}$ lie within the unit circle, the pair $(\mathcal{A}, \mathcal{B})$ is controllable, and the gain matrix $\mathcal{K}$ in \eqref{eq:discrete-feedback} that asymptotically stabilizes $\mathcal{O}^*$ is obtained using LQR with the weighting matrices $ \mathcal{Q} = \mathbb{I}_5$, $\mathcal{R} = 2\mathbb{I}_2$ as
\begin{equation*}
    \mathcal{K} = \begin{bmatrix}
    0.1048  & -0.0277  &  0.0267  &  0.0097  & -0.0000 \\
   -0.0048  &  0.0046  &  0.0037  &  0.0014  &  0.0000
    \end{bmatrix}
\end{equation*}

The same initial conditions \eqref{eq:initial-conditions} are used, and the simulation results are shown in Fig. \ref{fig:sim-orbit} for $k = 1$ through $k = 20$, a duration of approximately $10.49$ s.  The components of $\rho_k$ are shown in Fig. \ref{fig:sim-orbit}(a)-(b), and the components of $\drho_k$ are shown in Fig. \ref{fig:sim-orbit}(c)-(d). It is seen that the system trajectory converges to $\rho_k = 0$, and $\drho_k \to 0$ as $\rho_k \to 0$. The values of $\omega_k$, shown in Fig. \ref{fig:sim-orbit}(e), converge to values in agreement with \eqref{eq:omega-star-identical}, $\omega^* = -3.8483$ rad/s, indicating stabilization of the orbit in \eqref{eq:orbit-sim}. The time-of-flight, plotted in Fig. \ref{fig:sim-orbit}(f), converges to the constant value $\delta_k = 0.5442$ s. The control inputs $I_k$ and $r_k$ are shown in Fig. \ref{fig:sim-orbit}(g) and Fig. \ref{fig:sim-orbit}(h) respectively. The inputs are obtained from \eqref{eq:orbit-I-r} for $j = 1, 2, \dots, 7$, \emph{i.e.}, $k = 1,3, \dots, 13$. The controller $u(j)$ is inactive for $j > 7$ as the system trajectory is sufficiently close to $\mathcal{O}^*$. The inputs converge to $I_k = 0.7926$ Ns, $r_k = 0.0202$ m for $k$ odd, and $I_k = -0.4217$ Ns, $r_k = 0.0380$ m for $k$ even in accordance with \eqref{eq:Ik-stable} and \eqref{eq:rk-stable}. The trajectory of the center-of-mass of the devil-stick corresponding to these results is shown in Fig. \ref{fig:trajectory}(b).


\section{Conclusion}

The problem of juggling a devil-stick between two angular configurations was addressed using DVHCs. This paper extends the work in \cite{khandelwal_discrete_2025}, which considered \emph{rotations} of the passive coordinate to realize propeller motion, to \emph{oscillations} of the passive coordinate to realize planar juggling. 
The DVHC approach results in a rich set of juggling motions that are not necessarily symmetric about the vertical axis; the results for planar symmetric juggling in \cite{kant_non-prehensile_2021} are a special case. It is shown that DVHCs are an effective tool for design and control of the trajectory of an object juggled using impacts. 
Similar to the zero dynamics induced by VHCs, the DZD can be used to infer the nonlinear stability characteristics of the hybrid system subject to DVHCs, without relying on linearization about the fixed point of a Poincar\'e map \cite{kant_juggling_2022}.
Future work will focus on control design for the robot manipulator considering unilateral constraints and impact laws \cite{zavala-rio_control_1999, brogliato_control_2000}, robustness analysis of the system, and extension of the DVHC approach to general underactuated systems controlled using impulsive inputs.\

\newpage
\appendix
\begin{center}
    \vspace{-4pt}
    \textsc{Juggling between Symmetric Orientations}
\end{center}

\subsection{Enforcing the DVHC}

For the same physical parameters of the devil-stick, we present simulation results for the special case when $\thetaeven = \pi - \thetaodd$, \emph{i.e.}, the devil-stick is juggled between orientations symmetric about the vertical.
We consider the same DVHC as in \eqref{eq:vhc-sim}, with the values of $\thetaodd$ and $\thetaeven$ chosen to be
\begin{equation} \label{eq:theta-sim-symmetric}
    \thetaodd = \pi/6,\,\, \thetaeven = 5\pi/6 \ \Rightarrow\ \delths = 2\pi/3,\,\, \sigma = 0
\end{equation}

\noindent Again, the choices in \eqref{eq:vhc-sim} and \eqref{eq:theta-sim-symmetric} ensure that the DZD in \eqref{eq:zerodyn} permits stable, 2-periodic orbits - see Theorem \ref{theorem:periodic} and Corollary \ref{cor1}. The control inputs are obtained using \eqref{eq:Ik} and \eqref{eq:rk} with $\lambda$ in \eqref{eq:lambda-sim}.

\begin{figure}[t]
    \centering
    \psfrag{A}{\hspace{-10pt} \footnotesize{$\theta_k$ (rad)}}
    \psfrag{B}{\hspace{-18pt} \footnotesize{$\omega_k$ (rad/s)}}
    \psfrag{T}{\footnotesize{$k$}}
    \psfrag{Q}{\hspace{-10pt} \footnotesize{$\delta_k$ (s)}}
    \psfrag{F}{\hspace{-10pt} \footnotesize{$I_k$ (Ns)}}
    \psfrag{R}{\hspace{-10pt} \footnotesize{$r_k$ (m)}}
    \psfrag{C}{$\rho_x$}
    \psfrag{D}{$\rho_y$}
    \psfrag{G}{$\drho_x$}
    \psfrag{L}{$\drho_y$}
    \includegraphics[width=\linewidth]{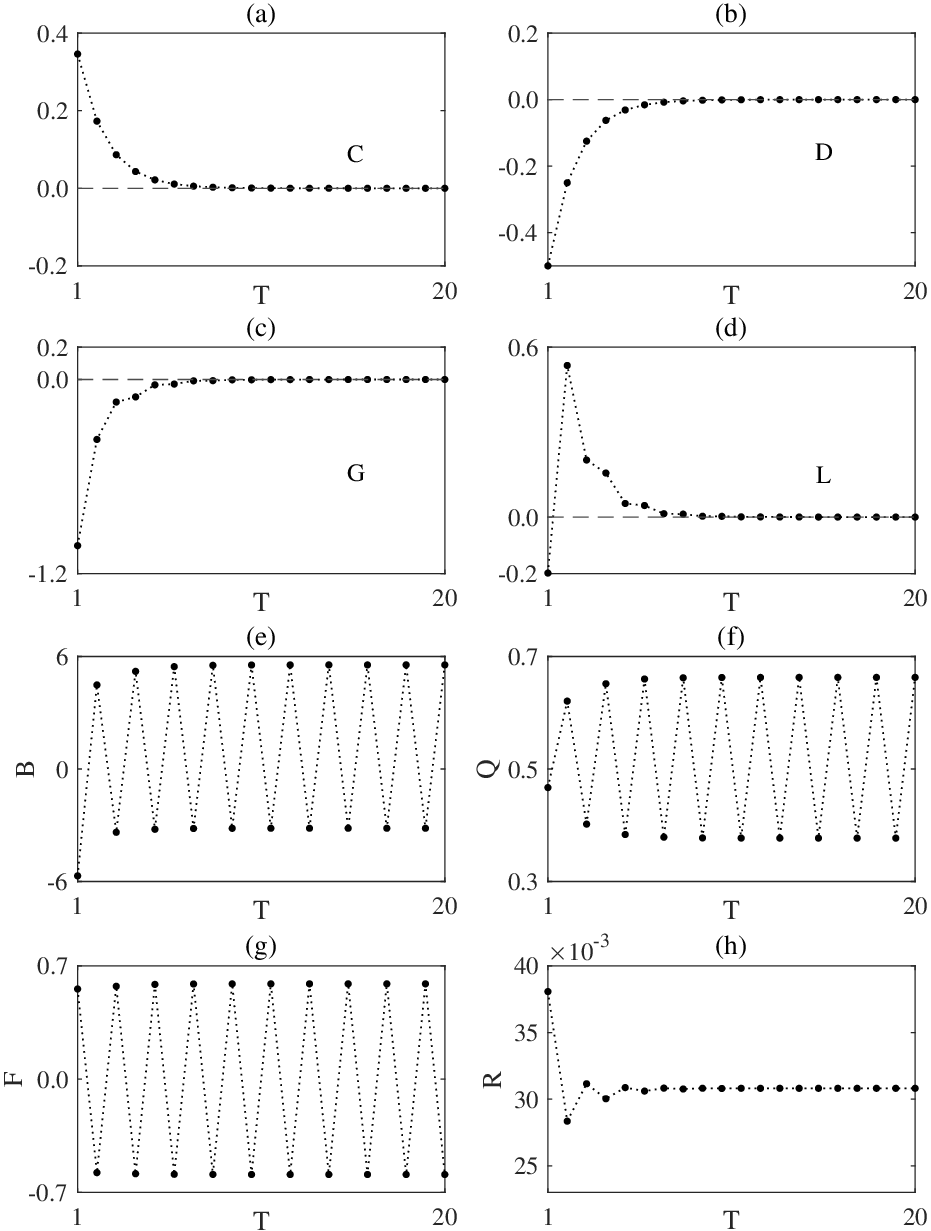}
    \caption{Stabilization of 2-periodic juggling of a devil-stick between orientations symmetric about the vertical from arbitrary initial conditions: 
    (a)-(b) show the components of $\rho_k$, (c)-(d) show the components of $\drho_k$,
    (e) shows the pre-impact angular velocity $\omega_k$, (f) shows the time-of-flight $\delta_k$
    (g) shows the applied impulse $I_k$, and (h) shows the point of application $r_k$ of the impulsive force.}
    \label{fig:sim-vhc-symmetric}
\end{figure}

We consider the initial conditions
\begin{equation} \label{eq:initial-conditions-symmetric}
    \begin{bmatrix} q^T & \dot q^T \end{bmatrix}^T =
    \begin{bmatrix} 0.7 & 2.5 & 0.5236 & 0.9 & -2.0 & -5.7 \end{bmatrix}^T 
\end{equation}

\noindent for which $\theta = \theta_1 = \thetaodd$ and $\rho_1 \neq 0$. The simulation results are shown in Fig. \ref{fig:sim-vhc-symmetric} for $k = 1$ through $k = 20$, corresponding to a duration of approximately $9.80$ s. The components of $\rho_k$ are shown in Fig. \ref{fig:sim-vhc-symmetric}(a)-(b), and the components of $\drho_k$ are shown in Fig. \ref{fig:sim-vhc-symmetric}(c)-(d). The plots demonstrate that the system trajectory converges to $\rho_k = 0$ exponentially. As the DVHC is enforced, the system settles to a stable, 2-periodic juggling motion with 
\begin{equation} \label{eq:periodic-juggling-sim-symmetric}
    \omega_k = \begin{cases}
        -3.1596,\  k\,\,\mathrm{odd} \\
        \ \ 5.5532,\ k\,\,\mathrm{even}
    \end{cases}
    \!\!\!\!\!\!,\,
    \delta_k = \begin{cases}
        0.3771,\  k\,\,\mathrm{odd} \\
        0.6629,\ k\,\,\mathrm{even}
    \end{cases}
\end{equation}

\noindent which agree with \eqref{eq:omega-odd-even} and \eqref{eq:delta-k-odd-even}. The values of $\omega_k$ and $\delta_k$ are plotted in Fig. \ref{fig:sim-vhc-symmetric}(e) and Fig. \ref{fig:sim-vhc-symmetric}(f).
The control inputs $I_k$ and $r_k$ are shown in Fig. \ref{fig:sim-vhc-symmetric}(g) and Fig. \ref{fig:sim-vhc-symmetric}(h) respectively. For the 2-periodic juggling motion in \eqref{eq:periodic-juggling-sim-symmetric}, $I_k = - (-1)^k \times 0.5890$ Ns and $r_k = 0.0308$ m, in accordance with \eqref{eq:Ik-stable} and \eqref{eq:rk-stable} - see Remark \ref{rem:inputs-symmetric}.
\begin{remark} \label{rem:inputs-symmetric}
    For $\thetaeven = \pi - \thetaodd$, \eqref{eq:Ik-stable} indicates that the \emph{magnitude} of impulse applied at either symmetric orientation of the devil-stick is the same, with opposite signs, and \eqref{eq:rk-stable} implies that the location of application of the impulse is identical for either symmetric orientation of the devil-stick.
\end{remark}

\begin{figure}[b]
    \centering
    \psfrag{X}{\hspace{-10pt} \footnotesize{$h_x$ (m)}}
    \psfrag{Y}{\hspace{-10pt} \footnotesize{$h_y$ (m)}}
    \includegraphics[width=\linewidth]{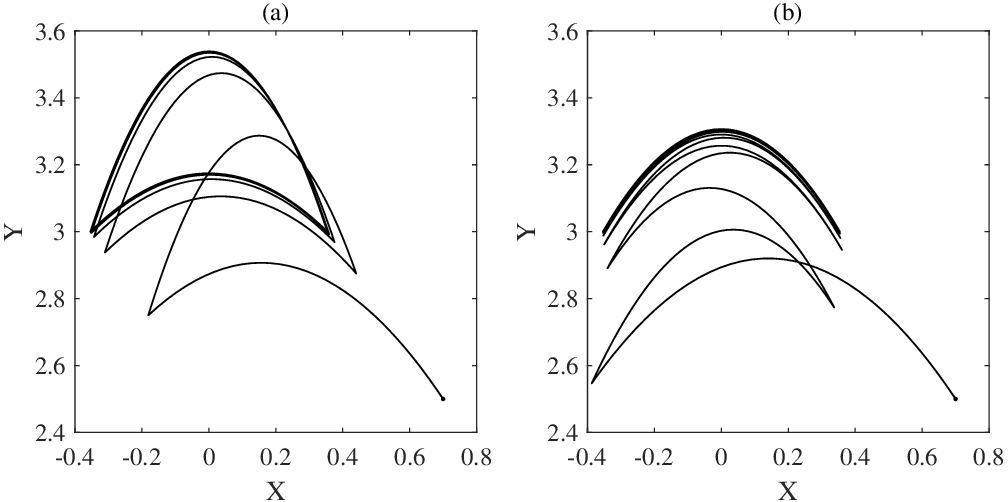}
    \caption{Trajectory of the center-of-mass of the devil-stick corresponding to the results in (a) Fig. \ref{fig:sim-vhc-symmetric} and (b) Fig. \ref{fig:sim-orbit-symmetric} showing planar symmetric juggling.}
    \label{fig:trajectory-symmetric}
\end{figure}

The trajectory of the center-of-mass of the devil-stick corresponding to these results is shown in Fig. \ref{fig:trajectory-symmetric}(a).

\subsection{Stabilization of a Periodic Orbit}

\begin{figure}[t]
    \centering
    \psfrag{A}{\hspace{-10pt} \footnotesize{$\theta_k$ (rad)}}
    \psfrag{B}{\hspace{-18pt} \footnotesize{$\omega_k$ (rad/s)}}
    \psfrag{T}{\footnotesize{$k$}}
    \psfrag{Q}{\hspace{-10pt} \footnotesize{$\delta_k$ (s)}}
    \psfrag{F}{\hspace{-10pt} \footnotesize{$I_k$ (Ns)}}
    \psfrag{R}{\hspace{-10pt} \footnotesize{$r_k$ (m)}}
    \psfrag{C}{$\rho_x$}
    \psfrag{D}{$\rho_y$}
    \psfrag{G}{$\drho_x$}
    \psfrag{L}{$\drho_y$}
    \includegraphics[width=\linewidth]{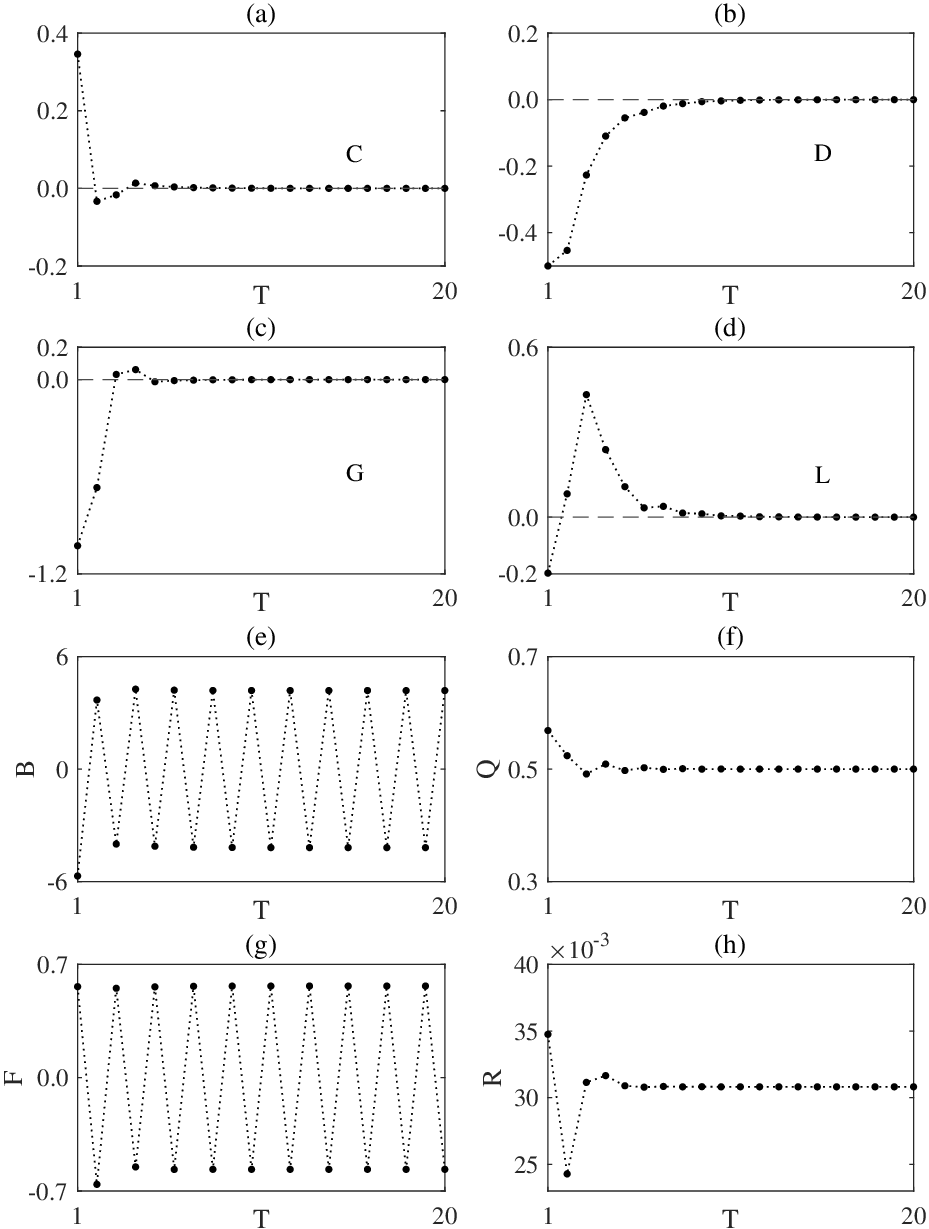}
    \caption{Orbital stabilization of planar symmetric juggling of a devil-stick from arbitrary initial conditions: 
    (a)-(b) show the components of $\rho_k$, (c)-(d) show the components of $\drho_k$,
    (e) shows the pre-impact angular velocity $\omega_k$, (f) shows the time-of-flight $\delta_k$
    (g) shows the applied impulse $I_k$, and (h) shows the point of application $r_k$ of the impulsive force.}
    \label{fig:sim-orbit-symmetric}
\end{figure}

We now demonstrate the effectiveness of the controller described in Section \ref{sec5} in stabilizing a desired orbit, specifically the orbit corresponding to planar symmetric juggling in \cite{kant_non-prehensile_2021, kant_juggling_2022}. This juggling motion is symmetric in all states about the vertical axis.
We consider the DVHC defined by \eqref{eq:vhc-sim} and \eqref{eq:theta-sim-symmetric}. Of the infinitely many stable, 2-periodic orbits, we choose to stabilize the orbit for which $\omega^* = -4.1888$ rad/s, obtained from \eqref{eq:omega-star-identical}. This results in $\delta_k = 0.5$ s $\forall k$ corresponding to stable juggling.
This orbit can be described using \eqref{eq:orbit} as
\begin{equation} \label{eq:orbit-sim-symmetric}
\begin{split}
    \mathcal{O}^* = \{(q, \dot q) : \qac(k) &= \Phi(\theta_k), \dqac(k) = \Psi(\theta_k, \omega_k), \\
    \omega_k &= -4.1888 \,\,\forall\, \theta_k = \pi/6 \}
\end{split}
\end{equation}

\noindent We obtain the control inputs \eqref{eq:Ik} and \eqref{eq:rk} with the same choice of $\lambda$ as in \eqref{eq:lambda-sim}. The section \eqref{eq:poincare-section} on which the orbit stabilizing controller \eqref{eq:orbit-I-r} acts is
\begin{equation} \label{eq:poincare-section-sim-symmetric}
    \Sigma = \{(q, \dot q) \in \mathbb{R}^2 \times S^1 \times \mathbb{R}^3 : \theta = \pi/6, \omega < 0\}
\end{equation}

\noindent The fixed point of the map $\mathbb{P}$ in \eqref{eq:fixed-point} corresponding to the orbit $\mathcal{O}^*$ in \eqref{eq:orbit-sim-symmetric} and the section $\Sigma$ in \eqref{eq:poincare-section-sim-symmetric} is
\begin{align}
    z^* &= \begin{bmatrix}
        0.3540 & 3.0000 & 1.4160 & -2.4525 & -4.1888
    \end{bmatrix}^T \notag \\
    I^* &= 0.5664, \quad r^* = 0.0308
\end{align}

\noindent The matrices $\mathcal{A}$ and $\mathcal{B}$ in \eqref{eq:linearized-map} are obtained numerically as
\begin{equation*}
\begin{split}
    \mathcal{A} &= \begin{bmatrix}
    0.2500  &  0.0000  &  0.0000  & -0.0000  & -0.0000 \\
    0.0000  &  0.2500  &  0.0000  &  0.0000  &  0.0000 \\
   -0.2500  &  0.4329  &  0.5001  &  0.2887  &  0.0000 \\
    0.4331  &  0.2495  &  0.8659  &  0.4999  &  0.0000 \\
   -0.7398  & -1.2807  & -1.4794  & -0.8541  & -0.0000
    \end{bmatrix}
    \\ 
    \mathcal{B} &= \begin{bmatrix}
    -0.0000  &  4.2847 & -12.2873 & -30.0787  & 36.3492 \\
   20.3351 &  31.1748 & -111.2006 & -200.6851 & 221.3658
    \end{bmatrix}^T
\end{split}
\end{equation*}

\noindent While not all eigenvalues of $\mathcal{A}$ lie within the unit circle, the pair $(\mathcal{A}, \mathcal{B})$ is controllable, and the gain matrix $\mathcal{K}$ in \eqref{eq:discrete-feedback} that asymptotically stabilizes $\mathcal{O}^*$ is obtained using LQR with the weighting matrices $ \mathcal{Q} = \mathbb{I}_5$, $\mathcal{R} = 2\mathbb{I}_2$ as
\begin{equation*}
    \mathcal{K} = \begin{bmatrix}
    0.0961 &   0.0358 &   0.0398  &  0.0230  &  0.0000 \\
   -0.0124  & -0.0017  & -0.0006  & -0.0003  &  0.0000
    \end{bmatrix}
\end{equation*}

The same initial conditions \eqref{eq:initial-conditions-symmetric} are used, and the simulation results are shown in Fig. \ref{fig:sim-orbit-symmetric} for $k = 1$ through $k = 20$, a duration of approximately $9.59$ s. The components of $\rho_k$ are shown in Fig. \ref{fig:sim-orbit-symmetric}(a)-(b), and the components of $\drho_k$ are shown in Fig. \ref{fig:sim-orbit-symmetric}(c)-(d). It is seen that the system trajectory converges to $\rho_k = 0$, and $\drho_k \to 0$ as $\rho_k \to 0$. 
The values of $\omega_k$, shown in Fig. \ref{fig:sim-orbit-symmetric}(e), converge to values in agreement with \eqref{eq:omega-star-identical}, $\omega^* = -4.1888$ rad/s, indicating stabilization of the orbit in \eqref{eq:orbit-sim-symmetric}. The time-of-flight, plotted in Fig. \ref{fig:sim-orbit-symmetric}(f), converges to the constant value $\delta_k = 0.5$ s. 
The control inputs $I_k$ and $r_k$ are shown in Fig. \ref{fig:sim-orbit-symmetric}(g) and Fig. \ref{fig:sim-orbit-symmetric}(h) respectively. The inputs are obtained from \eqref{eq:orbit-I-r} for $j = 1, 2, \dots, 7$, \emph{i.e.}, $k = 1,3, \dots, 13$. The controller $u(j)$ is inactive for $j > 7$ as the system trajectory is sufficiently close to $\mathcal{O}^*$. For planar symmetric juggling, $I_k = - (-1)^k \times 0.5664$ Ns and $r_k = 0.0308$ m in accordance with \eqref{eq:Ik-stable} and \eqref{eq:rk-stable} - see Remark \ref{rem:inputs-symmetric}. The trajectory of the center-of-mass of the devil-stick corresponding to these results is shown in Fig. \ref{fig:trajectory-symmetric}(b).

\balance
\bibliographystyle{IEEEtran}      
\bibliography{ref}   

\end{document}